\documentclass[12pt,preprint]{aastex}
\usepackage{amsmath}
\usepackage{graphicx}
\usepackage{color}
\usepackage{url}
\usepackage{CJKutf8}
\usepackage{ulem}

\slugcomment{Accepted by \aj}

\shorttitle{Archival C/2017 K2}
\shortauthors{Hui et al.}

\begin{document}

\title{
Prediscovery Observations and Orbit of Comet C/2017 K2 (PANSTARRS)
}
\author{
\begin{CJK}{UTF8}{bsmi}
Man-To Hui (許文韜)$^{1}$, 
\end{CJK}
David Jewitt$^{1,2}$
and David Clark$^{3,4}$
}
\affil{
$^1$Department of Earth, Planetary and Space Sciences,
UCLA, 595 Charles Young Drive East, Box 951567, 
Los Angeles, CA 90095-1567\\
}
\affil{
$^2$Department of Physics and Astronomy, UCLA, 
430 Portola Plaza, Box 951547, Los Angeles, CA 90095-1547\\
}
\affil{
$^3$Department of Physics and Astronomy, 
The University of Western Ontario,
London, Ontario, N6A 3K7, Canada\\
}
\affil{
$^4$Department of Earth Sciences, 
The University of Western Ontario,
London, Ontario, N6A 5B7, Canada\\
}
\email{pachacoti@ucla.edu}

\begin{abstract}

We present a study of comet C/2017 K2 (PANSTARRS) using prediscovery archival data taken from 2013 to 2017. Our measurements show that the comet has been marginally increasing in activity since at least 2013 May (heliocentric distance of $r_{\mathrm{H}} = 23.7$ AU pre-perihelion). We estimate the mass-loss rate during the period 2013--2017 as $\overline{\dot{M}} \approx \left(2.4 \pm 1.1 \right) \times 10^{2}$ kg s$^{-1}$, which requires a minimum active surface area of $\sim$10--10$^2$ km$^{2}$ for sublimation of supervolatiles such as CO and CO$_2$, by assuming a nominal cometary albedo $p_V = 0.04 \pm 0.02$. The corresponding lower limit to the nucleus radius is a few kilometers. Our Monte Carlo dust simulations show that dust grains in the coma are $\gtrsim0.5$ mm in radius, with ejection speeds from $\sim$1--3 m s$^{-1}$, and have been emitted in a protracted manner since 2013, confirming estimates by Jewitt et al.~(2017). The current heliocentric orbit is hyperbolic. Our N-body backward dynamical integration of the orbit suggests that the comet is most likely (with a probability of $\sim$98\%) from the Oort spike. The calculated median reciprocal of the semimajor axis  1 Myr ago was $a_{\mathrm{med}}^{-1} = \left( 3.61 \pm 1.71 \right) \times 10^{-5}$ AU$^{-1}$ (in a reference system of the solar-system barycentre).

\end{abstract}


\section{\uppercase{Introduction}}


Most comets are observed to show activity when they reach heliocentric distances $r_\mathrm{H} \lesssim 5$--6 AU, where the most abundant cometary volatile, water ice, begins to sublimate as a result of increasing insolation. However, a few comets have been observed to exhibit activity at greater heliocentric distances. This distant activity cannot be explained by sublimation of water ice, but other mechanisms including crystallization of amorphous ice (Prialnik \& Bar-Nun 1992) and sublimation of supervolatile species (A'Hearn et al.~2012) may be responsible. The reason why only a few distant comets have been observed to be active is twofold. Firstly, when comets are far away from the Sun, they are intrinsically less active because of their lower temperatures. Secondly, distant comets tend to be extremely faint, introducing an observational bias. 

Fortunately, thanks to ever-advancing technology and better sky coverage by ongoing sky surveys, recent years have witnessed an increasing number of discoveries of distant comets (e.g., C/2006 S3 (LONEOS) discovered at $r_{\mathrm{H}} = 14.3$ AU, C/2010 U3 (Boattini) at $r_{\mathrm{H}} = 18.4$ AU, both pre-perihelion), making a better understanding of activity in distant comets possible.

Comet C/2017 K2 (hereafter ``K2")  was detected by Pan-STARRS at Haleakala, Hawai`i on UT 2017 May 21 (Wainscoat et al.~2017), when it was $r_\mathrm{H} \approx 16$ AU from the Sun. The current orbital solution by the JPL Horizons ephemeris service identifies it as a long-period comet, with perihelion distance $q = 1.811$ AU, eccentricity $e = 1.00034$, inclination $i = 87\degr.6$, and a perihelion passage on UT 2022 Dec 21.\footnote{The elements are at epoch TT 2017 Jun 10.0, retrieved on 2017 Oct 04.} 

In earlier work (Jewitt et al.~2017), we used the Hubble Space Telescope (\textit{HST}) to set a limit to the size of the nucleus ($R_{\mathrm{N}} \lesssim 9$ km) and established that the coma of K2 consists of large ($\gtrsim 0.1$ mm) sized dust grains released over a period of years. We identified a prediscovery detection from 2013 at heliocentric distance $r_{\mathrm{H}} = 23.7$ AU. Meech et al.~(2017) argued instead that the coma grains are small ($\sim$1 $\mu$m) and, using a sublimation model, inferred a nucleus radius $14 \le R_{\mathrm{N}} \le 80$ km. Both papers conclude that the activity is likely driven by the sublimation of a supervolatile ice (CO, CO$_2$, N$_2$, or O$_2$ according to Jewitt et al.~(2017) and CO according to Meech et al.~(2017)). Other, non-equilibrium processes are also possible.


In this paper, we present archival, serendipitous prediscovery observations of K2, and we explore the orbit of K2 using Monte Carlo simulations.

\section{\uppercase{Observations}}


We used the Solar System Object Image Search (SSOIS; Gwyn et al.~2012) of the Canadian Astronomy Data Centre (CADC), to find K2 in archival data. Ten $U$-band images were found, taken at the 3.6-m Canada-France-Hawaii Telescope (CFHT) on UT 2013 May 10, 12 and 13, using the MegaCam prime focus imager. The offsets from ephemerides by JPL Horizons and the Minor Planet Center (MPC) at the time of our prediscovery were enormous, $\sim$$+3\arcmin$ in RA and $-11\arcmin$ in declination for the former, and $\sim$$+1\arcmin$ in RA and $-3\arcmin$ in declination for the latter, but the rate of angular motion  was fully consistent with both sources. Detailed descriptions of the observations and the image of the comet are given in Jewitt et al.~(2017). Judging from its non-stellar appearance ($\mathrm{FWHM} = 1\arcsec.5 \pm 0\arcsec.1$, compared to $\mathrm{FWHM} \approx 0\arcsec.9$ for nearby background stars), the comet was active in 2013, although it is slightly trailed in the data (0\arcsec.98 in length) because of the long exposure and its non-sidereal motion. 

Unfortunately none of the available star catalogs provide $U$-band magnitude data for stars in the field of view (FOV) of the CFHT data. Therefore, we calibrated $U$-band magnitudes of field stars then adjacent to the comet using the Keck-I 10-m telescope.  Observations on UT 2017 September 20 were taken using a $u'$-broadband filter (central wavelength 3404 \AA, full width at half maximum transmission 3750 \AA) under photometric conditions with the Low Resolution Imaging Spectrometer (LRIS; Oke et al.~1995).  The image scale was 0\arcsec.135 pixel$^{-1}$.  We calibrated field stars in the 2013 CFHT field using observations of photometric standard stars PG1648+536E and PG1633+099A  from the catalog by Landolt (1992).

\begin{figure}
\epsscale{1.0}
\begin{center}
\plotone{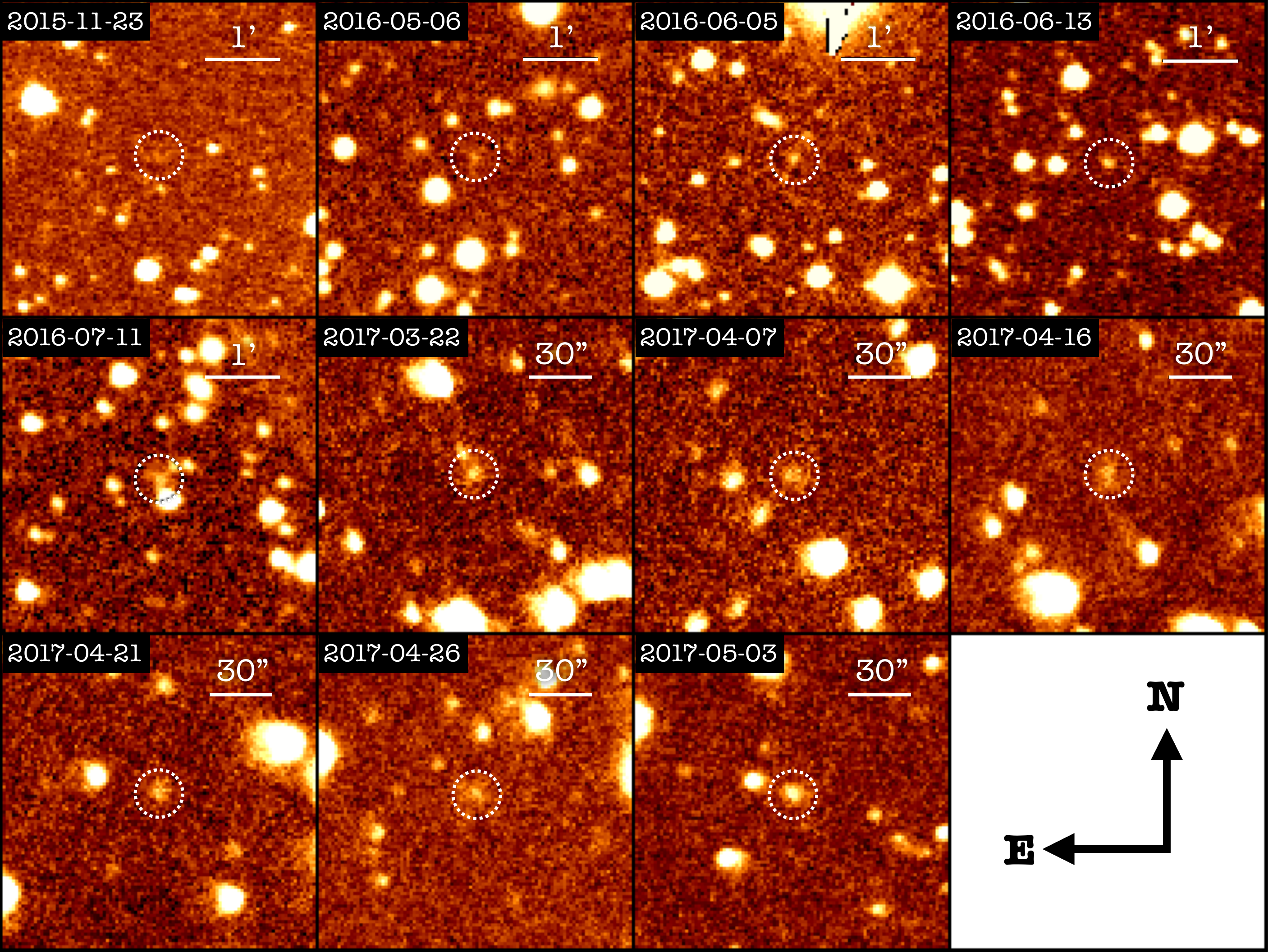}
\caption{
Composite prediscovery Catalina Sky Survey images of comet C/2017 K2 (PANSTARRS), with scale bars and observation dates labelled in each panel. The images are average coaddition from four-pass images taken at the same night with alignment on the comet, which is marked by a dashed circle in each of the panels. Equatorial north is up and east is left. We do not show the antisolar or the projected heliocentric velocity directions to avoid cluttering the plots. 
\label{fig_K2_CSS}
} 
\end{center} 
\end{figure}

Using image search software documented in Clark (2014), we also identified K2 in archival images from the Catalina Sky Survey (CSS) taken as early as 2015 November (see Table \ref{tab_geo} for details, and Figure \ref{fig_K2_CSS} for images), although with barely detectable motion with respect to the background sources, because of low angular resolution. The detections were consolidated by checking red plates from the Second Digitized Sky Survey (DSS2-red) having a similar limiting magnitude but a much better resolution. The CSS images from 2015 and 2016 were taken through a 0.7-m f/1.8 Schmidt telescope equipped with an unfiltered 4K $\times$ 4K CCD having an image scale of 2\arcsec.50 pixel$^{-1}$ and a FOV of 8.1 deg$^{2}$. Those from 2017 were obtained through the same telescope but equipped with an unfiltered 11K $\times$ 11K CCD having a scale of 1\arcsec.50 pixel$^{-1}$ and a FOV of 19.4 deg$^{2}$. All the images use an individual exposure time of $t_\mathrm{exp} = 30$ s, except that the set from 2015 has $t_\mathrm{exp} = 45$ s. The point-source $V$-band limiting magnitude of the images is $\sim$20.


We also searched for the comet in images taken between UT 2014 May 12 and 15 by the Palomar Transient Factory (PTF) 1.2-m diameter telescope (Law et al.~2009). The images, with an exposure time of 60 s and a FOV of $0\degr.6 \times 1\degr.1$, have a point-source $V$-band limiting magnitude of $\sim$20.5 (Waszczak et al.~2017), and a scale of 1\arcsec.01 pixel$^{-1}$. We had no success in detecting the comet in the individual images. Neither could we see anything above the noise level of the background in a stacked image coadded from the consecutive four-days of data with registration on the calculated motion of the comet. More recent PTF archival data from 2015 and 2016 cover the region of the comet, however, they are still proprietary, and we have no access to them. Given what we found with the CSS archival data, the comet should be detectable therein.

Prediscovery observations by the Pan-STARRS survey were reported as early as UT 2014 Mar 20 (Meech et al.~2017). We made no attempts to identify K2 in archival images prior to the CFHT observations in 2013, because the SSOIS inquiry shows no serendipitous observations from sufficiently large telescopes or with sufficiently long  exposure times.

\section{\uppercase{Results}}
\label{sec_rslt}

We performed aperture photometry on all the available images. Using an aperture 2\arcsec.3 in radius, we obtained the apparent magnitude of K2 in a stacked image, coadded from all the CFHT individual frames from UT 2013 May 10, 12 and 13 with registration on the apparent motion of the comet, as $m_U = 22.83 \pm 0.08$. This value benefits from the Keck calibration data described above and supercedes the coarse estimate ($m_U = 23.7 \pm 0.3$) by Jewitt et al.~(2017).

\begin{figure}
\epsscale{1.0}
\begin{center}
\plotone{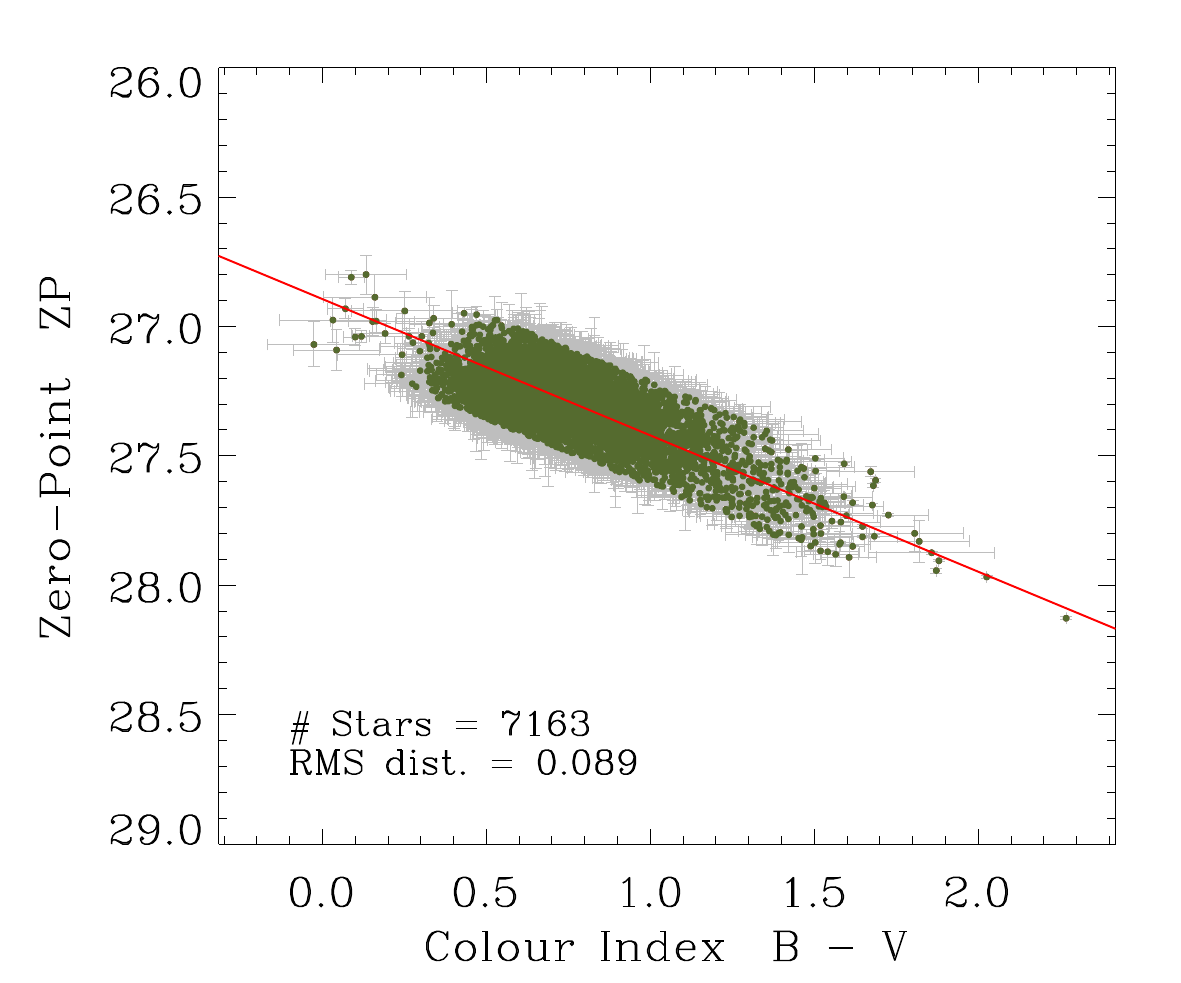}
\caption{
The zero-point of a coadded Catalina Sky Survey image from UT 2017 April 21 as a function of the color index of stars. The number of stars used is shown in the lower left of the plot. Also shown is the rms distance of the datapoints from the least-squares fit, which is drawn as the red line. Stars with residuals over $\pm$0.2 mag ($\sim$17\% of the total number) are discarded. This criterion has little effect on the derived zero-point (change by $\sim$0.02\%) and color term (change by $\sim$2.5\%).
\label{fig_CSS_ZP}
} 
\end{center} 
\end{figure}

We also performed aperture photometry on the CSS images. Since they were obtained without a photometric-standard filter, we had to first determine the zero-points of the images ($ZP$) by introducing $k_\mathrm{c}$, which is the color term satisfying the following equation

\begin{equation}
m_{\ast,V} - ZP_\lambda + 2.5 \log \left( \frac{\mathcal{F}_{\ast}}{t_\mathrm{exp}} \right) - k_\mathrm{c} \left( m_{\ast,B} - m_{\ast,V} \right) = 0
\label{eq_zp},
\end{equation}

\noindent where $m_{\ast, \lambda}$ is the star magnitude in some bandpass, and $\mathcal{F}_{\ast}$ is the signal in ADU of the star measured within photometric apertures of 15\arcsec.0 and 12\arcsec.0 in radius, respectively for the images taken by the old and new CCDs, which are approximately twice the FWHM of the field stars. The sky background was computed in annuli having inner and outer radii of $\sim$3$\times$ and $5 \times$ FWHM for the data, respectively. We utilised the AAVSO Photometric All-Sky Survey Data Release 9 (APASS-DR9; Henden et al.~2016) and reduced the zero-points of all the CSS images in the $V$ band from the least-squares fit (see Figure \ref{fig_CSS_ZP} as an example). The color of the comet ($m_U - m_V = 1.11 \pm 0.03$, $m_B - m_V = 0.74 \pm 0.02$, $m_V - m_R = 0.45 \pm 0.02$, from our Keck observation), is assumed to be unchanged. Then the measured flux of comet K2, which was obtained by applying the same photometric aperture and sky annulus that we used for stars, is converted to the apparent $V$-band magnitude. The errors stem mainly from the uncertainty in the determination of the zero-points ($\sim$0.2 mag), as well as the low signal-to-noise ratio of the comet.



\begin{figure}
  \centering
  \begin{tabular}[b]{@{}p{1\textwidth}@{}}
    \centering\includegraphics[scale=0.7]{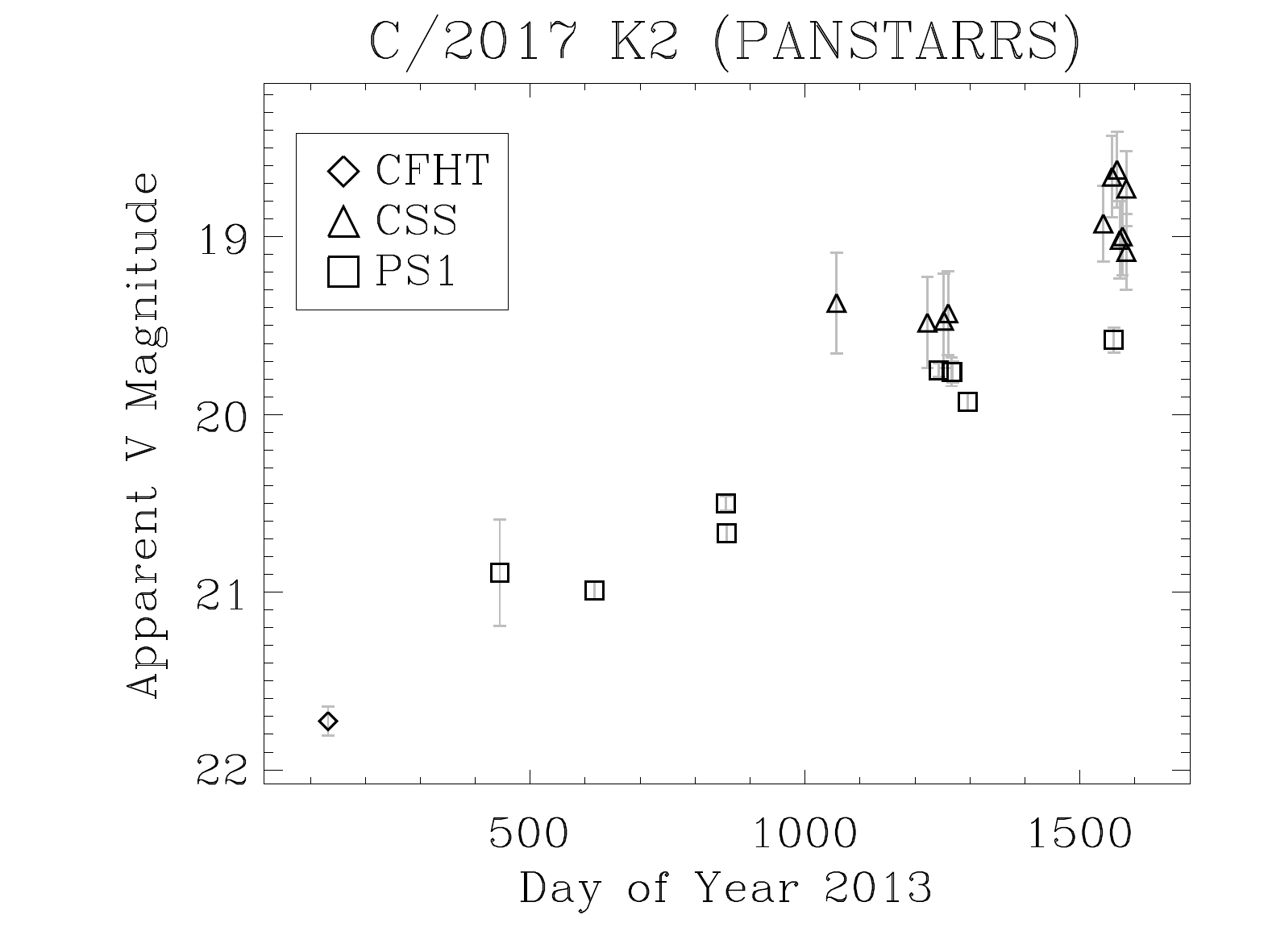} \\
    \centering\small (a)
  \end{tabular}%
  \quad
  \begin{tabular}[b]{@{}p{1\textwidth}@{}}
    \centering\includegraphics[scale=0.7]{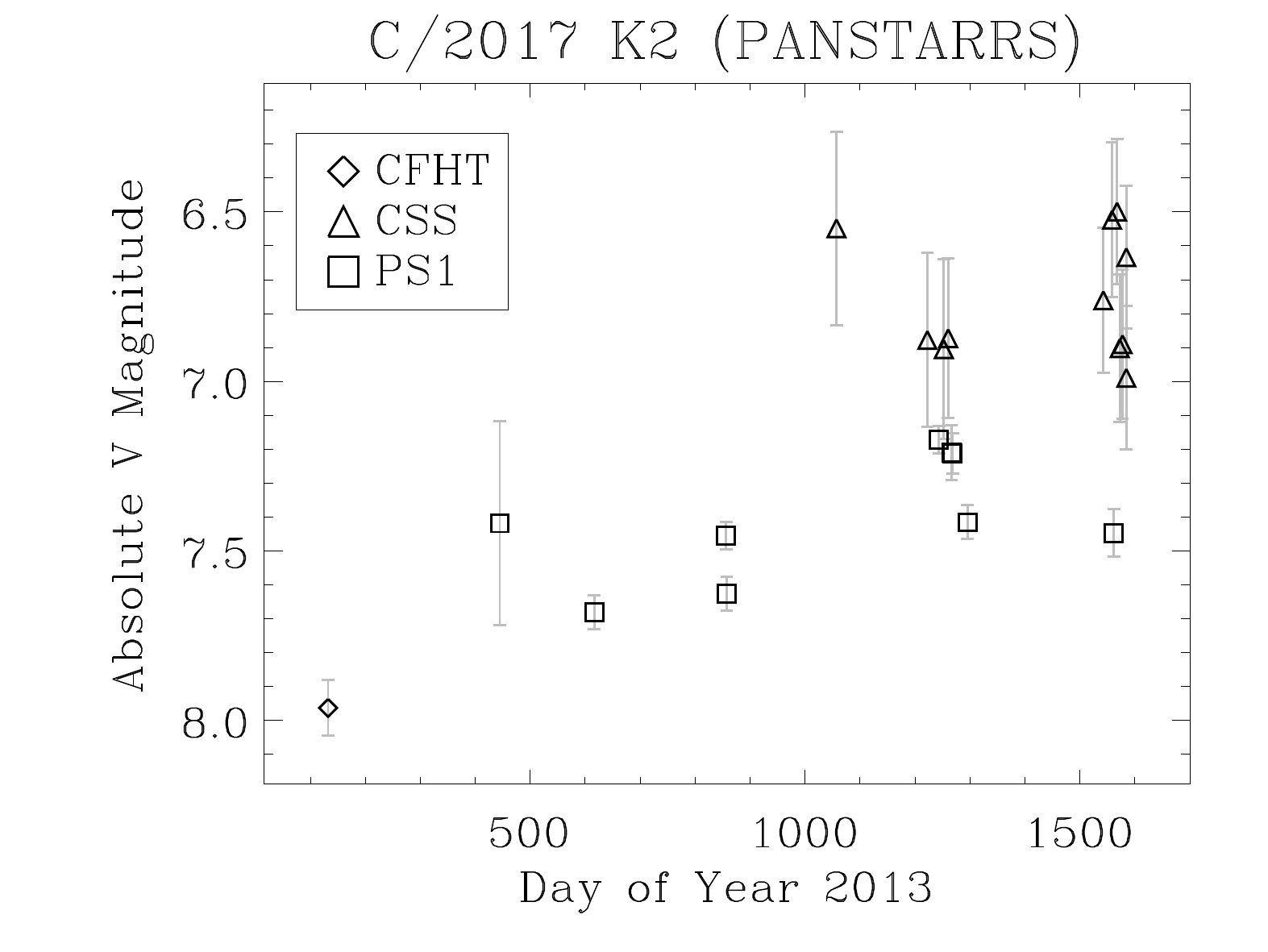} \\
    \centering\small (b)
  \end{tabular}
  \caption{
Temporal evolution of $V$-band magnitude of comet C/2017 K2 (PANSTARRS). PS1 refers to Pan-STARRS, whose datapoints are taken and converted from Meech et al.~(2017). Point symbols correspond to telescopes as shown in the legend. Panel (b) has apparent magnitude in panel (a) corrected to $r_{\mathrm{H}} = \mathit{\Delta} = 1$ AU and $\alpha = 0$\degr~from Equation (\ref{eq_H}).
  \label{fig_lc_2017K2}
  }
\end{figure}

Our results are summarised in Table \ref{tab_mag}. The temporal evolution of the apparent magnitude of K2 is shown in Figure \ref{fig_lc_2017K2}(a), in which we have included photometry from Meech et al.~(2017).\footnote{Magnitude data by Meech et al.~(2017) from the prediscovery Pan-STARRS observations were converted from Sloan-$r'$ to $V$-band magnitude using a transformation equation from Jordi et al.~(2006).} Note that different sizes of photometric apertures have been employed (see Table \ref{tab_mag}). Meech et al.~(2017) scaled their measurements by means of curves of growth to a set that would have been obtained using an aperture of 5\arcsec.0 in radius, which is much smaller than aperture sizes we applied to the CSS images. As the comet is a diffuse source, $\sim$9\arcsec~in radius in June 2017 (Jewitt et al.~2017), we expect that their measurements  systematically underestimate the brightness (as is evident in Figure \ref{fig_lc_2017K2}). The closer the comet, the more significant is the difference. For example, for a steady-state coma with sufficient signal-to-noise ratio, the aperture correction in the prediscovery Pan-STARRS observations from $\mathit{\Delta} = 22.2$ to 16.3 AU results in a difference of $\sim$0.3 mag, with the more recent photometry being too faint. Indeed, this phenomenon can be readily seen in Figure \ref{fig_lc_2017K2}(a).  However, our attempts to correct the Pan-STARRS photometry to an aperture of fixed linear (as opposed to angular) radius  based on the surface brightness profile obtained in Jewitt et al.~(2017) failed to give satisfactory results.   We simply decided not to perform aperture corrections.

\begin{figure}
\epsscale{1.0}
\begin{center}
\plotone{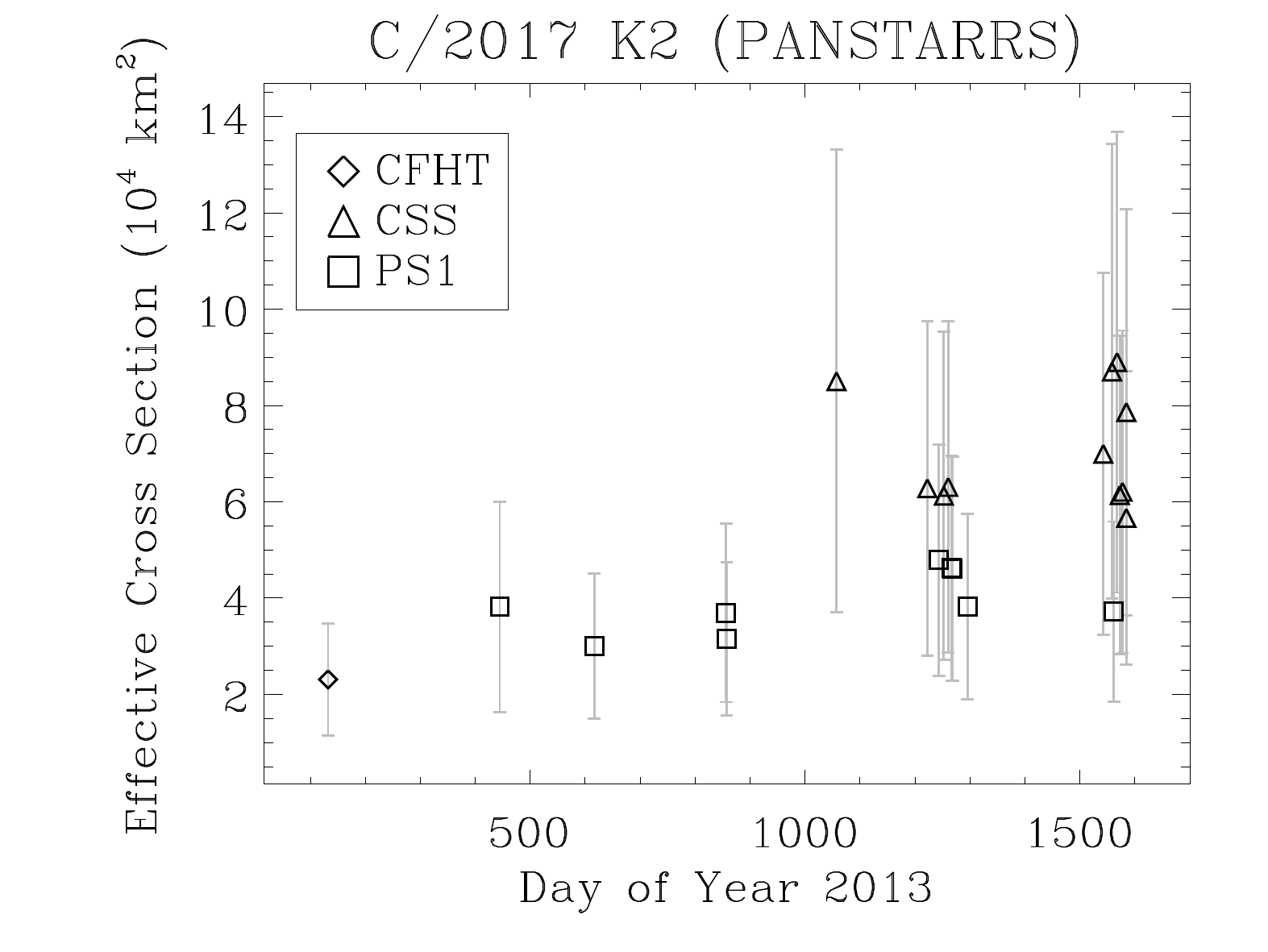}
\caption{
The effective cross-section of comet C/2017 K2 (PANSTARRS) as a function of time. Datapoints from Pan-STARRS are taken and computed from Meech et al.~(2017). Point symbols correspond to telescopes as shown in the legend. The errors are propagated from uncertainties in the magnitude data and the error in albedo.
\label{fig_xs_2017K2}
}
\end{center}
\end{figure}

In order to investigate the intrinsic brightness of the comet, the effect of the varying viewing geometry ought to be eliminated, so we compute the absolute magnitude from

\begin{equation}
m_V \left(1,1,0 \right) = m_V \left( r_\mathrm{H}, \mathit{\Delta}, \alpha \right) - 5 \log \left(r_\mathrm{H} \mathit{\Delta} \right) + 2.5 \log \phi \left( \alpha \right)
\label{eq_H},
\end{equation}

\noindent where $\alpha$ is the phase angle, and $\phi (\alpha)$ is the phase function of the coma.  For the latter, we use the empirical phase function of dust (Marcus 2007; Schleicher \& Bair 2011; \url{http://asteroid.lowell.edu/comet/dustphase.html}), and normalised at $\alpha=0$\degr. The result is shown in Figure \ref{fig_lc_2017K2}(b). We then estimate the effective cross-section, $C_\mathrm{e}$, using

\begin{equation}
C_\mathrm{e} = \frac{\pi r_{\oplus}^2}{p_{V}} 10^{-0.4 \left[m_{V} \left(1,1,0 \right) - m_{\odot,V}\right]}
\label{eq_Xsect},
\end{equation}

\noindent where $r_\oplus \approx 1.5 \times 10^{8}$ km is the mean Sun-Earth distance, $p_{V}$ is the $V$-band geometric albedo, assumed to be $p_V = 0.04 \pm 0.02$ (see Lamy et al.~2004), and $m_{\odot,V} = -26.74$ is the $V$-band apparent magnitude of the Sun. We plot the temporal variation of the effective cross-section of K2 in Figure \ref{fig_xs_2017K2}, and summarise our photometry in Table \ref{tab_mag}.

\section{\uppercase{Discussion}}

\subsection{Dust Dynamics}
\label{sec_dd}

We attempted to constrain properties of dust grains of comet K2 in the high-resolution post-discovery \textit{HST} observation by Jewitt et al.~(2017) using our Monte Carlo model, in which dust grains are released from the nucleus with a range of non-zero initial velocities, and are subsequently subject to the solar radiation pressure force and the gravitational force from the Sun, whose ratio is denoted as $\beta$. The grain size is related to $\beta$ by $\beta \propto \left(\rho_{\mathrm{d}} \mathfrak{a} \right)^{-1}$, where $\mathfrak{a}$ is dust grain size, and $\rho_{\mathrm{d}}$ is bulk density. Together with the release time of the dust particles from the observed epoch $\tau$, their trajectories can be uniquely determined. We assumed that the dust particles follow a simple power-law size distribution, i.e., $\mathrm{d} n \left(\mathfrak{a} \right) \propto \mathfrak{a}^{-3.5} \mathrm{d}\mathfrak{a}$, where $\mathrm{d} n $ is the number of dust grains having radii between $\mathfrak{a}$ and $\mathfrak{a} + \mathrm{d} \mathfrak{a}$, and that their ejection speeds are described by an empirical relationship $v_{\mathrm{ej}} \propto \mathfrak{a}^{-0.5}$. Similar Monte Carlo models have been widely applied elsewhere (e.g., Fulle 1989; Ishiguro 2008; Moreno 2009; Ye \& Hui 2014; Hui et al.~2017).

\begin{figure}
\epsscale{1.0}
\begin{center}
\plotone{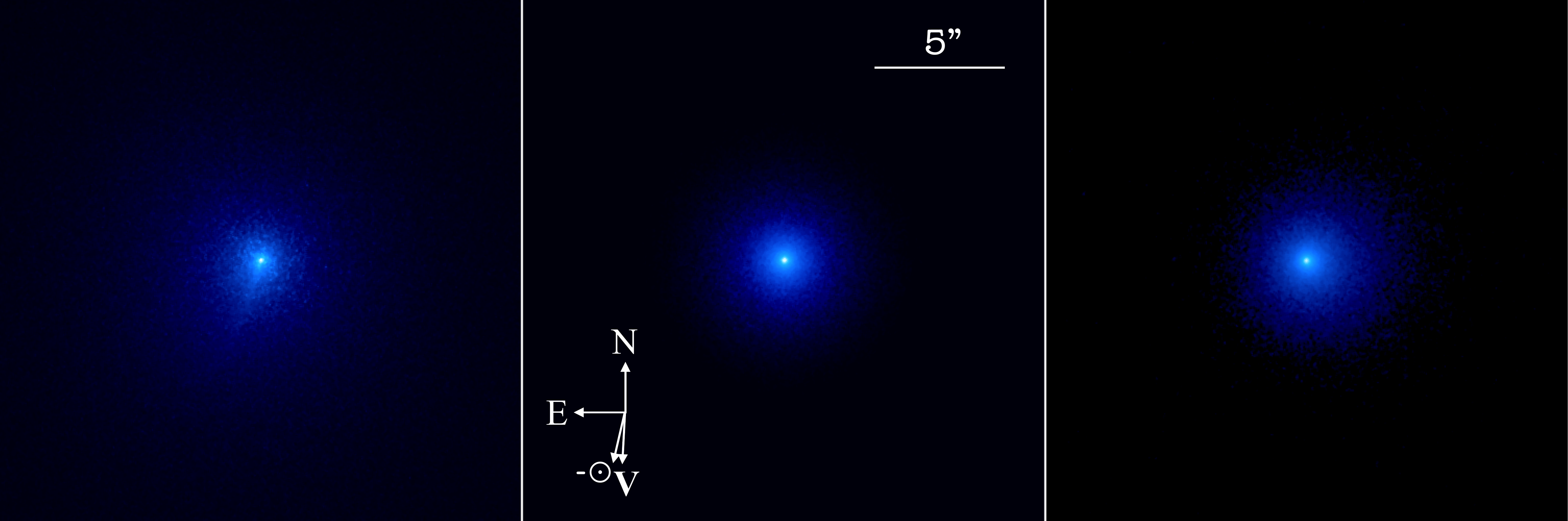}
\caption{
Monte Carlo models of comet C/2017 K2 (PANSTARRS) comparing size distributions with minimum particle radius (left) $\mathfrak{a}_{\min}$ = 10 $\mu$m and (middle) $\mathfrak{a}_{\min}$ = 500 $\mu$m. Dust in both models is assumed to follow a power-law distribution of radii with index $-3.5$ and to extend up to largest radius $\mathfrak{a}_{\max}$ = 2 mm. Ejection speeds for dust grains of 1 mm for the left two panels are both 1.9 m s$^{-1}$. The large-particle model (middle panel) closely matches the nearly circular coma in \textit{HST} data from UT 2017 June 27 (right panel, also see Figure 1 in Jewitt et al.~(2017) for isotopes).  The small-particle model (left panel) shows a prominent tail which is not present in the data. The initial dust release time is set to $\tau_{0} = 1500$ days. A total number of $\sim$10$^{6}$--10$^{7}$ particles were generated in both simulations. Dimensions of each panel are $20\arcsec \times 20\arcsec$. The cardinal directions and the projected antisolar ($-\odot$) direction and the heliocentric velocity vector ($\mathbf{V}$) are indicated. 
\label{fig_mdl_17K2}
} 
\end{center} 
\end{figure}

We tested our simulation with different combinations of minimum, $\mathfrak{a}_{\min}$, and maximum, $\mathfrak{a}_{\max}$, particle radius,   earliest dust release time from the observed epoch, $\tau_{0}$, and $v_{\mathrm{ej}}$. A successful model should be able to match the observed morphology of the coma. The model is insensitive to the size of the largest dust grains, $\mathfrak{a}_{\max}$, because such particles are rare and carry a negligible fraction of the total scattering cross-section for the power-law index assumed. In Figure (\ref{fig_mdl_17K2}) we show two models computed using  $\mathfrak{a}_{\max} = 2$ mm, $v_{\mathrm{ej}}$ = 1.9 m s$^{-1}$ and $\tau_{0} = 1500$ days, consistent with the CFHT observation that K2 has been active at least since 2013 May. The remaining dust parameters can then be obtained without much ambiguity. By inspection, we determine that large dust grain (radii of $0.5 \lesssim \mathfrak{a} \lesssim 2$ mm) models  closely simulate the morphology of comet K2 in the \textit{HST} data (c.f.~the middle- and right-hand panels of Figure \ref{fig_mdl_17K2}). On the other hand, models including smaller particles (radii of $0.01 \lesssim \mathfrak{a} \lesssim 2$ mm) show a clear radiation-pressure-swept tail which is not present in the data (c.f.~left and right-hand panels of the Figure).  The Monte Carlo models thus support the inference made by Jewitt et al.~(2017) to the effect that the coma is dominated by submillimeter particles, but contradicts the one by Meech et al.~(2017), who assumed 2 $\mu$m-sized dust particles in their sublimation model. We found that this conclusion cannot be mitigated even if  more recent ejection times, $\tau_{0}$, and higher ejection speeds are adopted. To conclude, the observed dust particles of comet K2 must be large, at least submillimeter sized, to avoid the formation of an observable, radiation pressure swept tail.

\subsection{Mass Loss}
\label{sec_ml}

There is a large scatter in the absolute magnitudes (see Figure \ref{fig_lc_2017K2}(b)) in part due to the aperture issue discussed in Section (\ref{sec_rslt}).  Nevertheless, the comet appears to brighten in data from 2017 compared to 2013. The brightening corresponds to a maximum possible increase in the scattering cross-section $\Delta C_\mathrm{e} \approx \left(4.5 \pm 2.1 \right) \times 10^{4}$ km$^{2}$.  The mean mass-loss rate of the comet, denoted as $\overline{\dot{M}}$, can be estimated from

\begin{equation}
\overline{\dot{M}} = \frac{4\rho_{\mathrm{d}} \Delta C_\mathrm{e} \bar{\mathfrak{a}}}{3\Delta t}
\label{eq_mloss},
\end{equation}

\noindent where $\rho_{\mathrm{d}}$ and $\bar{\mathfrak{a}}$ are the bulk density and the mean radius of the dust grains, respectively. With a nominal $\rho_{\mathrm{d}} = 0.5$ g cm$^{-3}$,  $\bar{\mathfrak{a}} \approx 1$ mm, and $\Delta t \approx 1.2 \times 10^{8}$ s, Equation (\ref{eq_mloss}) yields $\overline{\dot{M}} \approx \left(2.4 \pm 1.1 \right) \times 10^{2} $ kg s$^{-1}$. The uncertainty only incorporates the error from photometry together with the error in albedo. Our estimate for the mass-loss rate is larger than $\overline{\dot{M}} \sim 60$ kg s$^{-1}$ by Jewitt et al.~(2017), mainly because we adopted a larger mean grain size based on our Monte Carlo simulations. Given the fact that the maximum dust dimension cannot be confidently constrained and that there are many other unknowns, such a difference is not significant.

A lower limit to the size of the nucleus of comet K2 can be estimated by assuming that the activity is supported by equilibrium sublimation of exposed ices. We solve the energy equilibrium equation between insolation, thermal emission and sublimation,

\begin{equation}
\left(1 - A_{\mathrm{B}} \right) S_{\odot}\left( \frac{r_{\oplus}}{r_{\mathrm{H}}} \right)^{2} \cos \zeta = \epsilon \sigma T^{4} + L\left( T \right) f_{\mathrm{s}} \left( T \right)
\label{eq_sub},
\end{equation}

\noindent in which $A_{\mathrm{B}}$ is the Bond albedo, $S_{\odot} = 1361$ W m$^{-2}$ is the solar constant, $\cos \zeta$ is the effective projection factor for the surface ($\cos \zeta = 1$ for a subsolar scenario, and $\cos \zeta = 1/4$ corresponding to an isothermal nucleus), $\epsilon$ is the emissivity, $\sigma = 5.67 \times 10^{-8}$ W m$^{-2}$ K$^{-4}$ is the Stefan-Boltzmann constant, $T$ is the surface temperature in K, and $L$ (in J kg$^{-1}$) and $f_{\mathrm{s}}$ (in kg m$^{-2}$ s$^{-1}$) are the latent heat of the sublimating material, and the mass flux of the sublimated ice, respectively, both as functions of temperature. The heat conduction towards the nucleus interior is ignored. For simplicity, we assign $\epsilon = 0.9$ and $A_{\mathrm{B}} = 0.01$ (e.g., Buratti et al.~2004) and analyse the sublimation of CO (whose volatility is representative of other potential supervolatiles like N$_2$ and O$_2$) and CO$_{2}$ at the time-averaged heliocentric distance $r_{\mathrm{H}} = 17.1$ AU during the period 2013--2017.

We adopted empirical thermodynamic parameters of CO and CO$_2$ respectively listed in Prialnik et al.~(2004) and Cowan \& A'Hearn (1979), solved Equation (\ref{eq_sub}) and obtained $4.9 \times 10^{-6} \le f_{\mathrm{s}} \le 2.0 \times 10^{-5}$ kg m$^{-2}$ s$^{-1}$ for CO, and $8.6 \times 10^{-11} \le f_{\mathrm{s}} \le 2.2 \times 10^{-6}$ kg m$^{-2}$ s$^{-1}$ for CO$_{2}$, where the lower limits correspond to isothermal sublimation and the upper ones are from subsolar sublimation. In order to supply the mass-loss rate inferred from photometry, the minimum surface area, $A_{\mathrm{s}} = \overline{\dot{M}} / f_{\mathrm{s}}$, has to be in the range $12 \lesssim A_{\mathrm{s}} \lesssim 48$ km$^2$ for sublimation of CO, and $1.1 \times 10^{2} \lesssim A_{\mathrm{s}} \lesssim 2.8 \times 10^{6}$ km$^2$ for CO$_{2}$. These are equivalent to equal-area circles of radii $R_{\mathrm{N}} \sim \sqrt{A_{\mathrm{s}} / \pi} \gtrsim 2$ km and $R_{\mathrm{N}} \gtrsim 6$ km, respectively.  Given that the upper limit to the nucleus radius from the HST measurement is $R_{\mathrm{N}} <$ 9 km, we see that sublimation of CO (and N$_2$, O$_2$) is  easily capable of supplying the coma even if only a small fraction of the nucleus surface is active, while CO$_2$ must be sublimating from near the subsolar point, if it is present.

We then proceed to estimate the critical grain size, $\mathfrak{a}_{\mathrm{c}}$, of dust particles, which can be lifted off from the surface by the gas-drag force $F_{\mathrm{D}} = C_{\mathrm{D}} \pi \mathfrak{a}^2 \mu \mathfrak{m}_{\mathrm{H}} N v_{\mathrm{th}}^{2}$, where $C_{\mathrm{D}}$ is the dimensionless drag coefficient, $\mu$ is the molecular weight ($\mu = 28$ for CO, and $\mu = 44$ for CO$_2$), $\mathfrak{m}_{\mathrm{H}} = 1.67 \times 10^{-27}$ kg is the mass of the hydrogen atom, $N$ is the number density of the molecule, and $v_{\mathrm{th}}$ is the thermal speed of the gas. By equating the gas-drag force and the gravitational force at the surface and ignoring spinning of the body, with simple algebra we derive the critical grain dimension as

\begin{equation}
\mathfrak{a}_{\mathrm{c}} = \frac{9C_{\mathrm{D}} \overline{\dot{M}}}{32 \pi^2 G \rho \rho_{\mathrm{d}} R^3} \sqrt{\frac{2 k_{\mathrm{B}} T}{\pi \mu \mathfrak{m}_{\mathrm{H}}}}
\label{eq_ac},
\end{equation}

\noindent where $G = 6.67 \times 10^{-11}$ m$^{3}$ kg$^{-1}$ s$^{-2}$ is the gravitational constant, $\rho$ is the density of the nucleus, and $k_{\mathrm{B}} = 1.38 \times 10^{-23}$ J K$^{-1}$ is the Boltzmann constant. Assuming a unity gas-to-dust production ratio, along with $C_{\mathrm{D}} = 1$ and $\rho = \rho_{\mathrm{d}}$, Equation (\ref{eq_ac}) yields $\mathfrak{a}_{\mathrm{c}} \lesssim 4$ mm for sublimation of CO, and $\mathfrak{a}_{\mathrm{c}} \lesssim 0.2$ mm for CO$_2$, which is in line with Jewitt et al.~(2017). Note that our dust model suggests the size of the dust grains $\mathfrak{a} \gtrsim 0.5$ mm. We thus prefer CO (and materials of similar volatility) sublimation as the cause of the activity, but since there are many approximations in our model (e.g.~the neglect of rotation and the neglect of contact forces at the nucleus surface, see Gundlach et al.~2015), we feel that it would be premature to rule out CO$_2$ as the activity driver.

\subsection{Orbital Evolution}
\label{sec_orb}

We next examine the dynamical evolution of comet K2 in an attempt to understand its recent history. We downloaded the astrometric measurements of the comet from the MPC, which include our own astrometry from the prediscovery archival images. The measurements were debiased following Farnocchia et al.~(2015). The code \textit{EXORB9}, a part of the \textit{SOLEX12} package developed by A. Vitagliano, was exploited for orbit determination. Weights on each set of observations were adjusted to approximately accommodate ad hoc astrometric residuals whenever they were found aggressive. Twenty-four observations with residuals greater than 1\arcsec.5 either in RA or declination were discarded, leaving 336 observations (93\% of the total number) to be fitted by orbit determination. An optimised solution was thereby obtained, having a weighted rms of 0\arcsec.472. Our derived orbital elements are generally similar to those in the solution by JPL Horizons (Table \ref{tab_orb}), despite different choices of the weighting scheme and the threshold for filtering bad-residuals astrometry (D. Farnocchia, private communication). We then generated 10$^4$ clones of the nominal orbit according to the associated covariance matrix of the orbital elements, and performed backward N-body integration in \textit{MERCURY6} (Chambers 1999) using the 15th-order RADAU integrator (Everhart 1985) for the past 1 kyr, and the hybrid symplectic algorithm for the past 1 Myr to investigate the dynamical evolution of K2.\footnote{Switching to the hybrid integration scheme is a measure to reduce the computation time, at the cost of losing some accuracy. Since we have no interest in examining orbits of individual clones, but are looking at the total statistics, the result is not influenced.} Gravitational perturbations from the eight major planets and Pluto, post-Newtonian corrections (Arminjon 2002), and the influence of the galactic tide, which is a major perturber of the Oort cloud (e.g., Heisler \& Tremaine 1986; Fouchard et al.~2005), were included in the simulation. The possible distant giant planet claimed by Trujillo \& Sheppard (2015) and Batygin \& Brown (2016) was not considered, since the evidence for this body remains equivocal (Shankman et al.~2017). Neither have we included stellar perturbations, although the frequency of encounters with stellar systems passing within 1 pc of the Sun is estimated to be as many as $11.7 \pm 1.3$ Myr$^{-1}$ (Garc\'{i}a-S\'{a}nchez et al.~2001), yet 73\% of the encounters are with M dwarfs having low masses ($\lesssim 0.4 M_{\odot}$, where $M_{\odot}$ is the solar mass).

We did not incorporate possible non-gravitational acceleration of K2 in the orbital solution. To test the impact of this neglect, we employed \textit{EXORB9} to repeat the aforementioned procedures to solve for non-gravitational parameters $A_{j}$ ($j=1,2,3$) as defined in Marsden et al.~(1973) but obeying an empirical momentum-transfer law from sublimation of CO and CO$_2$ in a hemispherical scenario,  following the method in Hui \& Jewitt (2017). No detection of non-gravitational acceleration was made above the formal uncertainty levels (well below $1\sigma$). Solving for the non-gravitational parameters barely helps reduce the rms of the fit (to 0\arcsec.471),  justifying our omission of the non-gravitational effect.

\begin{figure}
\epsscale{1.0}
\begin{center}
\plotone{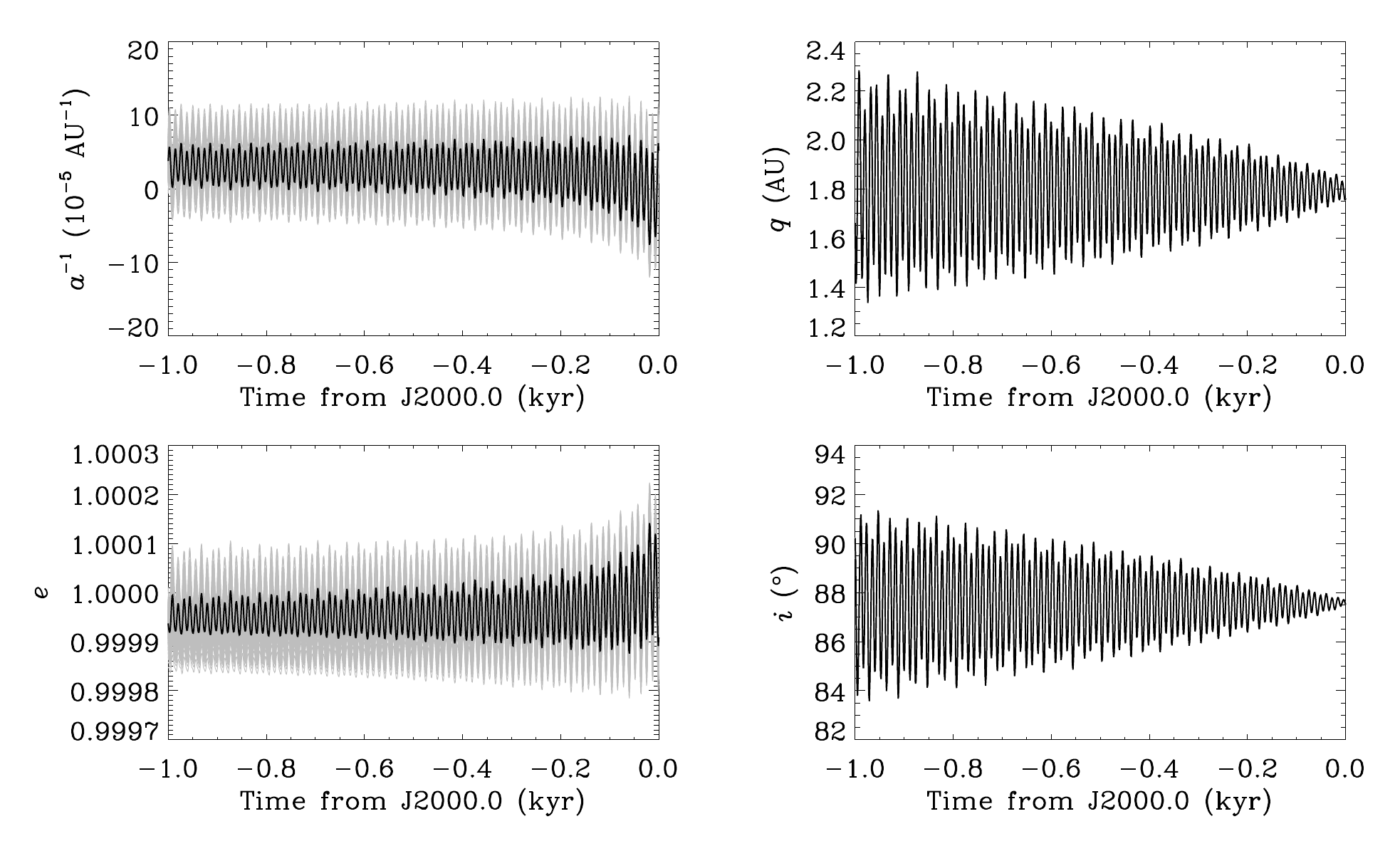}
\caption{
Orbital evolution of the nominal orbit (black) and 500 of the 10000 Monte Carlo clones (grey) of C/2017 K2 (PANSTARRS) in the past 1 kyr under the heliocentric reference system. In the right two panels about evolution of perihelion distance and inclination, all of the clones follow basically the same trends. Since the clones are synthesised from the nominal orbit, the median values of the four orbital elements as functions of time can be represented by the nominal orbit.
\label{fig_oscele_1ky_17K2}
} 
\end{center} 
\end{figure}

The orbital evolution of K2 in terms of the reciprocal of the semimajor axis ($a^{-1}$), perihelion distance $q$, eccentricity $e$ and inclination $i$ in the past 1 kyr is shown in Figure \ref{fig_oscele_1ky_17K2}. Note that the orbital elements are still referred to the heliocentric reference system. We can see that the ranges of $q$ and $i$ sway increasingly with time in the past 1 kyr, whereas $a^{-1}$ gradually approaches $\sim$10$^{-5}$ AU$^{-1}$ and $e$ tends to creep $< 1$. The examined orbital elements exhibit zigzagging oscillations with a dominant period of $\sim$11.9 yr, close enough to the orbital period of Jupiter to indicate non-negligible gravitational perturbations from the gas giant. 

\begin{figure}
\epsscale{1.0}
\begin{center}
\plotone{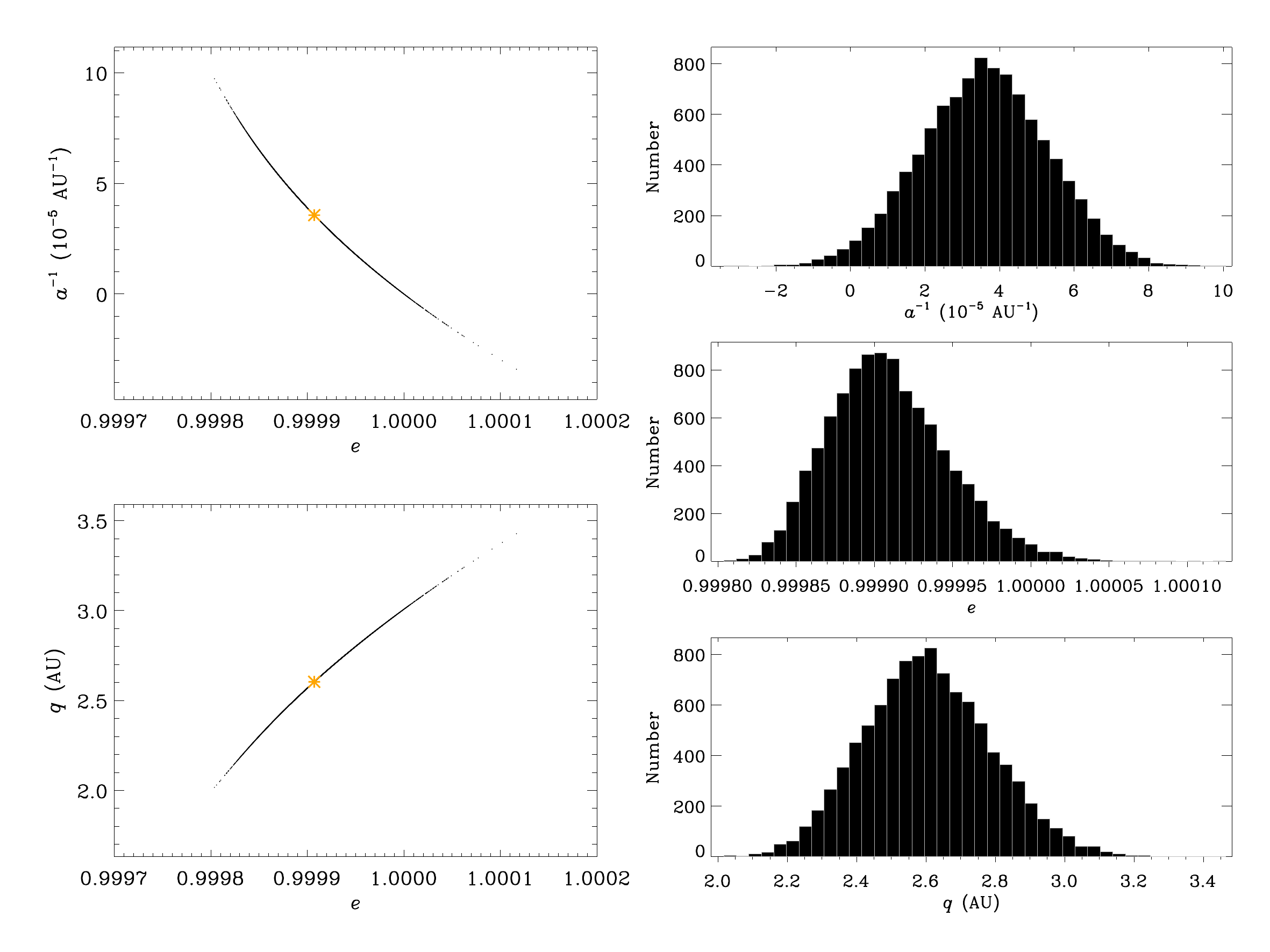}
\caption{
The left two panels show the past motion of the nominal orbit (orange asterism) and the 10000 Monte Carlo clones (black dots) of C/2017 K2 (PANSTARRS), in terms of distribution in the $e$-$a^{-1}$ and $e$-$q$ planes at 1 Myr ago from J2000.0. Also plotted are the histograms of orbital elements $a^{-1}$, $e$ and $q$ at the same epoch, in the right three panels. Note that the orbital elements here are referred to the solar-system barycentric reference system.
\label{fig_baryorb_1Myr_17K2}
} 
\end{center} 
\end{figure}

Now we move on to results from the backward integration for the past 1 Myr. Starting from now, we change the reference origin to the barycenter of the solar system. We obtain median values $a_{\mathrm{med}}^{-1} = \left(3.61 \pm 1.71 \right) \times 10^{-5}$ AU$^{-1}$ and $e_{\mathrm{med}} < 1$ from the clones (See Figure \ref{fig_baryorb_1Myr_17K2}, the assigned uncertainty is the standard deviation). Only 173 ($\sim$1.7\%) of the total clones have originally hyperbolic orbits and so we conclude that the comet is very unlikely to be of interstellar origin.  Instead, K2 is probably from the Oort spike, which consists of a mix of dynamically new and old comets (Kr{\'o}likowska \& Dybczy{\'n}ski 2010; Fouchard et al.~2013).  We cannot determine whether the comet is dynamically new or old from our backward integration, because the integration time (1 Myr) is shorter than the orbital period of the comet, $P = a^{3/2} \gtrsim 2$ Myr. Only from the region with $a^{-1} < 2.5 \times 10^{-6}$ AU$^{-1}$ in the Oort spike, are dynamically old comets completely absent (Kr{\'o}likowska \& Dybczy{\'n}ski 2017). As a result, whether K2 penetrated into the planetary region during the previous perihelion passage must be regarded as unsettled. Analysis of the forward integration of the orbit of K2 is not performed, because we have concern that intensified  sublimation activity as the comet approaches the Sun will intensify   non-gravitational effects.

Although detected as early as 2013, K2 managed to repeatedly escape detection by the major sky surveys. Why was comet K2 not discovered much earlier? The two important reasons, we suspect, are the high inclination and the low rate of its angular motion. The majority of sky surveys are optimised for making discoveries of small bodies that move at much higher speeds, such as near-Earth and main-belt asteroids. Furthermore, angualr resolution has been generally sacrificed for wider-FOV coverage, making discovery of slow-moving objects even more difficult. Although surveys like the Outer Solar System Origin Survey\footnote{\url{http://www.ossos-survey.org/about.html}} are dedicated to transneptunian objects, and should have had capability to detect objects moving as slowly as K2, they mainly search along the ecliptic plane.

\section{\uppercase{Summary}}

Key conclusions of our study about comet C/2017 K2 (PANSTARRS) are summarised as follows.

\begin{enumerate}

\item The comet was recorded serendipitously by the CFHT and the CSS on many occasions since 2013. At $r_{\mathrm{H}} = 23.7$ AU, K2 is the most distant comet ever observed on the way to perihelion.

\item The combined archival photometry suggests that the activity of the comet has been slowly increasing since 2013, as it approaches the Sun.

\item By means of our Monte Carlo simulation of the dust motion, we confirm that dust properties estimated by Jewitt et al.~(2017) during the \textit{HST} observation are valid, i.e., predominant dust grains of the comet are $\gtrsim 0.5$ mm in radius, with ejection speeds of $\sim$1--3 m s$^{-1}$, and have been released in a continuous manner since 2013 May. 

\item By assuming a cometary albedo $p_V = 0.04 \pm 0.02$, the mass-loss rate of comet K2 during the period of 2013--2017 was estimated to be $\overline{\dot{M}} \approx \left(2.4 \pm 1.1 \right) \times 10^{2}$ kg s$^{-1}$, which requires a minimum active surface area of $\sim$12 km$^{2}$ if the activity is driven by sublimation of CO, and $\sim$110 km$^{2}$ for CO$_2$. The nucleus must be at least of kilometer-size to sustain the observed activity by sublimation.

\item Monte Carlo simulations of the pre-entry orbit of the comet give original (1 Myr ago) reciprocal semimajor axis  $a_{\mathrm{med}}^{-1} = \left(3.61 \pm 1.71 \right) \times 10^{-5}$ AU$^{-1}$ (referred to the barycenter of the solar system).   We find  that some 98\% of orbital clones originate from within the Oort spike.  

\end{enumerate}

\acknowledgements
{
We thank Eric Christensen, Davide Farnocchia, Aldo Vitagliano, and Quan-Zhi Ye for their generous help, and the anonymous referee for a speedy review. This research is in part based on observations obtained with MegaPrime/MegaCam, a joint project of CFHT and CEA/DAPNIA, at the Canada-France-Hawaii Telescope (CFHT) which is operated by the National Research Council (NRC) of Canada, the Institut National des Sciences de l'Univers of the Centre National de la Recherche Scientifique of France, and the University of Hawaii. The facilities of the Canadian Astronomy Data Centre operated by the National Research Council of Canada with the support of the Canadian Space Agency, and the AAVSO Photometric All-Sky Survey (APASS), funded by the Robert Martin Ayers Sciences Fund, were used. Some of our data presented herein were obtained at the W. M. Keck Observatory, which is operated as a scientific partnership among the California Institute of Technology, the University of California and the National Aeronautics and Space Administration. The Observatory was made possible by the generous financial support of the W. M. Keck Foundation. The products of the Digitized Sky Surveys, which were produced at the Space Telescope Science Institute under U.S. Government grant NAG W-2166, were employed in this research. The NASA/ IPAC Infrared Science Archive, which is operated by the Jet Propulsion Laboratory, California Institute of Technology, under contract with the NASA, has been made use of. This work is funded by a grant from NASA to D.J.
}

\clearpage{}

\begin{deluxetable}{lcccccccc}
\tablecaption{Viewing Geometry of C/2017 K2 (PANSTARRS) in the Archival Data
\label{tab_geo}}
\tablewidth{0pt}
\tablehead{ 
\colhead{Date (UT)} & \colhead{Telescope\tablenotemark{[1]}} & \colhead{Filter} & \colhead{$r_\mathrm{H}$ (AU)\tablenotemark{[2]}}  & \colhead{$\mathit{\Delta}$ (AU)\tablenotemark{[3]}} & \colhead{$\alpha$ (\degr)\tablenotemark{[4]}}  & \colhead{$\varepsilon$ (\degr)\tablenotemark{[5]}}  & \colhead{$X$\tablenotemark{[6]}} & \colhead{Condition\tablenotemark{[7]}} 
}
\startdata

2013-05-10 & CFH & $U$ & 23.75 & 23.77 & 2.4 & 87.6 &  1.6 & A \\
2013-05-12 & CFH & $U$ & 23.74 & 23.77 & 2.4 & 87.5 &  1.6 &  \\
2013-05-13 & CFH & $U$ & 23.74 & 23.76 & 2.4 & 87.4 &  1.7 & A \\
2015-11-23 & CSS & -- & 19.09 & 19.16 & 3.0 & 84.2 &  1.8 & M \\
2016-05-06 & CSS & -- & 18.20 & 18.16 & 3.2 & 90.6 &  1.2& A \\
2016-06-05 & CSS & -- & 18.04 & 18.01 & 3.2 & 90.3 &  1.2& \\
2016-06-13 & CSS & -- & 18.00 & 17.97 & 3.2 & 90.0 &  1.3& \\
2016-07-11 & CSS & -- & 17.84 & 17.84 & 3.3 & 88.8 &  1.3& \\
2017-03-22 & CSS & -- & 16.43 & 16.41 & 3.5 & 89.3 &  1.2 & M \\
2017-04-07 & CSS & -- & 16.34 & 16.30 & 3.5 & 90.4 &  1.2& \\
2017-04-16 & CSS & -- & 16.29 & 16.24 & 3.5 & 91.0 &  1.2 & M \\
2017-04-21 & CSS & -- & 16.26 & 16.21 & 3.5 & 91.3 &  1.2 & M \\
2017-04-26 & CSS & -- & 16.23 & 16.17 & 3.6 & 91.5 &  1.2& \\
2017-05-03 & CSS & -- & 16.19 & 16.13 & 3.6 & 91.8 &  1.2& \\

\enddata

\tablenotetext{[1]}{CFH = the Canada-France-Hawaii Telescope; CSS = the Catalina Sky Survey}
\tablenotetext{[2]}{Heliocentric distance}
\tablenotetext{[3]}{Topocentric distance}
\tablenotetext{[4]}{Phase angle}
\tablenotetext{[5]}{Solar elongation}
\tablenotetext{[6]}{Air mass, dimensionless}
\tablenotetext{[7]}{A = astronomical twilight; M = moonlight}

\tablecomments{
The table excludes negative prediscovery observations of the comet.
}
\end{deluxetable}


\begin{deluxetable}{lcccccc}
\tablecaption{Archival Photometry of C/2017 K2 (PANSTARRS)
\label{tab_mag}}
\tablewidth{0pt}
\tablehead{ 
\colhead{Date (UT)} & \colhead{Telescope} & \colhead{$\vartheta$ (\arcsec)\tablenotemark{[1]}} & \colhead{$\lambda$\tablenotemark{[2]}} & \colhead{$m_\lambda \left(r_\mathrm{H}, \mathit{\Delta}, \alpha \right)$\tablenotemark{[3]}} & 
\colhead{$m_V \left(1, 1, 0 \right)$\tablenotemark{[4]}} &
\colhead{$C_\mathrm{e}$ (10$^4$ km)\tablenotemark{[5]}}
}
\startdata

2013-05-12\tablenotemark{\dagger} & CFH & 2\arcsec.3 & $U$ & $22.83 \pm 0.08$ & $7.96 \pm 0.08$ & $2.3 \pm 1.2$ \\
2015-11-23 & CSS & 15\arcsec.0 & $V$ & $19.37 \pm 0.29$ & $6.55 \pm 0.29$ & $8.5 \pm 4.8$ \\
2016-05-06 & CSS & 15\arcsec.0 & $V$ & $19.48 \pm 0.26$ & $6.88 \pm 0.26$ & $6.3 \pm 3.5$\\
2016-06-05 & CSS & 15\arcsec.0 & $V$ & $19.47 \pm 0.27$ & $6.91 \pm 0.27$ & $6.1 \pm 3.4$ \\
2016-06-13 & CSS & 15\arcsec.0 & $V$ & $19.43 \pm 0.24$ & $6.87 \pm 0.24$ & $6.3 \pm 3.4$ \\
2017-03-22 & CSS & 12\arcsec.0 & $V$ & $18.93 \pm 0.21$ & $6.76 \pm 0.21$ & $7.0 \pm 3.8$ \\
2017-04-07 & CSS & 12\arcsec.0 & $V$ & $18.66 \pm 0.23$ & $6.52 \pm 0.23$ & $8.7 \pm 4.7$ \\
2017-04-16 & CSS & 12\arcsec.0 & $V$ & $18.62 \pm 0.21$ & $6.50 \pm 0.21$ & $8.9 \pm 4.8$ \\
2017-04-21 & CSS & 12\arcsec.0 & $V$ & $19.02 \pm 0.22$ & $6.90 \pm 0.22$ & $6.1 \pm 3.3$ \\
2017-04-26 & CSS & 12\arcsec.0 & $V$ & $19.00 \pm 0.22$ & $6.89 \pm 0.22$ & $6.2 \pm 3.4$ \\
2017-05-03 & CSS & 12\arcsec.0 & $V$ & $18.73 \pm 0.21$ & $6.63 \pm 0.21$ & $7.9 \pm 4.2$ \\
2017-05-03 & CSS & 12\arcsec.0 & $V$ & $19.09 \pm 0.21$ & $6.99 \pm 0.21$ & $5.7 \pm 3.0$ \\

\enddata

\tablenotetext{[1]}{Aperture radius}
\tablenotetext{[2]}{Reduction filter}
\tablenotetext{[3]}{Apparent magnitude in the corresponding reduction filter}
\tablenotetext{[4]}{Absolute magnitude}
\tablenotetext{[5]}{Effective cross-section}
\tablenotetext{\dagger}{Mid-observation date for the CFHT data from UT 2013 May 10, 12 and 13}

\tablecomments{
Photometry for the CFHT prediscovery data was conducted on the image coadded from UT 2013 May 10, 12 and 13 with alignment on the apparent motion of comet C/2017 K2 (PANSTARRS). In images from UT 2016 July 11, the comet almost overlapped a background star, making photometry impossible.
}
\end{deluxetable}

\clearpage

\begin{deluxetable}{l|cccc}
\tabletypesize{\footnotesize}
\tablecaption{Orbital Elements
(Heliocentric Ecliptic J2000.0)
\label{tab_orb}}
\tablewidth{0pt}
\tablehead{ Orbital Element & 
\multicolumn{2}{|c}{This Work}  & 
\multicolumn{2}{c}{JPL Horizons\tablenotemark{\dagger}} \\
 & 
\colhead{Value} & \colhead{1$\sigma$ Uncertainty} & 
\colhead{Value} & \colhead{1$\sigma$ Uncertainty} 
}
\startdata
Perihelion distance $q$ (AU) 
       & 1.811198 & 1.08$\times$10$^{-4}$ 
       & 1.811103 & 1.63$\times$10$^{-4}$ \\ 
Orbital eccentricity $e$ 
       & 1.000350 & 3.10$\times$10$^{-5}$ 
       & 1.000337 & 5.16$\times$10$^{-5}$ \\ 
Orbit inclination $i$ (\degr) 
       & 87.55423 & 4.21$\times$10$^{-5}$ 
       & 87.55420 & 1.16$\times$10$^{-4}$ \\ 
Longitude of ascending node $\Omega$ (\degr)
                 & 88.17645 & 4.42$\times$10$^{-4}$ 
                 & 88.17642 & 1.21$\times$10$^{-3}$ \\ 
Argument of perihelion $\omega$ (\degr)
                 & 236.01570 & 3.60$\times$10$^{-3}$ 
                 & 236.01760 & 5.71$\times$10$^{-3}$ \\ 
Time of perihelion $t_\mathrm{p}$ (TT)
                  & 2022 Dec 21.394 & 7.37$\times$10$^{-2}$ 
                  & 2022 Dec 21.391 & 1.30$\times$10$^{-1}$ \\ 

\enddata
\tablenotetext{\dagger}{The solution was retrieved on 2017 Oct 04.}
\tablecomments{
Both solutions have orbital elements at a common epoch of JD 2457914.5 = 2017 June 10.0 TT. In our solution, the total number of astrometric observations, which span from UT 2013 May 10 to 2017 Sept 25, is 336. JPL Horizons used 351 observations covering an arc from UT 2013 May 12 to 2017 September 25. The weighted rms of our solution is $\pm0\arcsec.472$, whereas JPL Horizons only shows a dimensionless normalised rms of $\pm0.524$ for its solution.
}
\end{deluxetable}

\begin{figure}
\epsscale{0.7}
\begin{center}
\plotone{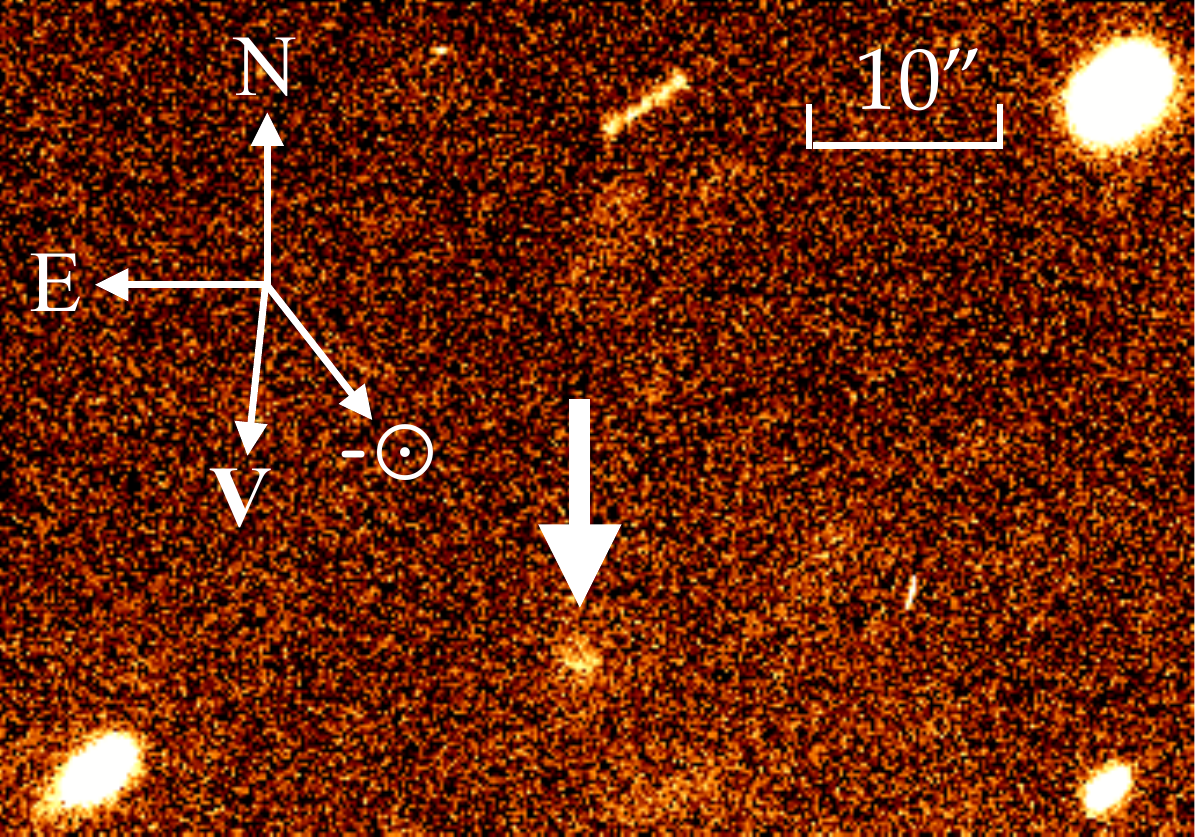}
\caption{
The CFHT average image of comet C/2017 K2 (PANSTARRS) coadded from ten individual $U$-band prediscovery images from UT 2013 May 10 to 13 with alignment on the motion of the comet. The effective total integration time is 100 min. As indicated by a compass in the upper left corner, equatorial north is up and east is left. The antisolar direction ($-\odot$) and the heliocentric velocity vector projected on the sky plane ($\mathbf{V}$), respectively, are both shown. Also shown is a scale bar. The image has angular dimensions $\sim$$1\arcmin.1 \times 0\arcmin.7$. The streaks are trails of background stars and galaxies.
\label{fig_K2_CFHT}
} 
\end{center} 
\end{figure}

\end{document}